\documentstyle[preprint,prb,aps,epsf,eqsecnum]{revtex}
\begin{document}
%\draft
\tightenlines

\title{Root of unity symmetries in the 8 and 6 vertex models}

\author{Klaus Fabricius
\footnote{e-mail Fabricius@theorie.physik.uni-wuppertal.de}}
\address{ Physics Department, University of Wuppertal, 
42097 Wuppertal, Germany}
\author{Barry~M.~McCoy
\footnote{e-mail mccoy@insti.physics.sunysb.edu}}               
\address{ Institute for Theoretical Physics, State University of New York,
 Stony Brook,  NY 11794-3840}
\date{\today}
\preprint{YITPSB-04-60}

\maketitle

\begin{abstract}

We review the recently discovered symmetries of the 8 and 6 vertex
models which exist at roots of unity and present their relation with
representation theory of affine Lie algebras, Drinfeld polynomials 
and Bethe vectors.
\end{abstract}

%\begin{flushright}
% {\tt YITP-SB-04-??}
%\end{flushright}

\medskip \noindent
{\bf Keywords:} lattice models, Bethe equations, quantum groups, loop
algebras, Drinfeld polynomial

\section{Introduction}
\label{intro}

The 8 vertex model is a lattice model in statistical mechanics
whose transfer matrix is given in ref. \cite{bax72} by
\begin{equation}
{\bf T}_8(v)|_{\mu,\nu}={\rm Tr}W(\mu_1,\nu_1)W(\mu_2,\nu_2)\cdots W(\mu_N,\nu_N)
\end{equation}
where $\mu_j,\nu_j=\pm1$ and $W(\mu,\nu)$ is a $2\times 2$ matrix whose
nonvanishing elements are given as
\begin{eqnarray}
W(1,1)|_{1,1}&=W(-1,-1)|_{-1,-1}=\rho\Theta(2\eta)\Theta(v-\eta)H(v+\eta)=a(v)
\nonumber\\
W(-1,-1)|_{1,1}&=W(1,1)|_{-1,-1}=\rho\Theta(2\eta)H(v-\eta)\Theta(v+\eta)=b(v)
\nonumber\\
W(-1,1)|_{1,-1}&=W(1,-1)|_{-1,1}=\rho H(2\eta)\Theta(v-\eta)\Theta(v+\eta)=c(v)
\nonumber\\
W(1,-1)|_{1,-1}&=W(-1,1)|_{-1,1}=\rho H(2\eta)H(v-\eta)H(v+\eta)=d(v).
\label{bw8}
\end{eqnarray}
The definition and some useful properties of $H(v)$ and $\Theta(v)$
are summarized in  appendix A.
This model is characterized by
the important property that for all fixed $\eta$ all elliptic nomes $p$ 
and all chain lengths $N$
it  satisfies the
commutation relation \cite{bax72}
\begin{equation}
[{\bf T}(v),{\bf T}(v')]=0.
\label{tt}
\end{equation}

The 8 vertex model also has the important property that there are
many cases in which the eigenvalues and eigenvectors of the transfer
matrix may be computed, but, in contrast with the commutation relation
(\ref{tt}), qualifying statements must be made on the allowed values of
$\eta$ and $N.$ Some of these qualifying statements are present in the
original studies of the eigenvalues \cite{bax72}, \cite{baxb} and
eigenvectors \cite{bax731}-\cite{bax733} by Baxter and others have
been recently observed by the present authors \cite{fm4}-\cite{fm5}

\vspace{.1in}

{\bf Conditions for eigenvalues}
 
1) $N$ unrestricted with $\eta=mK/L$ with $L$ even or $L$ and $m$ odd
 (ref. \cite{bax72})

2) $N$ even and $\eta$ unrestricted (ref.\cite{bax731}, \cite{baxb}) 

3) $\eta=mK/L$ with $L$ odd, $m$ even   
 and $N$ even with $N\leq L-1$ (ref. \cite{fm4}).

\vspace{.1in}

{\bf Conditions for some eigenvectors}

 1) $N$ even and $\eta=mK/L$ (ref. \cite{bax731}-\cite{bax733})

2) $N$ odd and $\eta=mK/L$ with $L$ odd, $m$ even and $N=2n_B+n_LL$ 
with $n_B$ and $n_L$ integers (ref.\cite{bax731}-\cite{bax733}).

\vspace{.1in}

This array of qualifying conditions is in contrast with the special
case of the 6 vertex model where the nome $p$ vanishes, the Boltzmann
weight $d(v)$ vanishes and the remaining nonvanishing weights are
\begin{eqnarray}
W(1,1)|_{1,1}&=W(-1,-1)|_{-1,-1}=\rho'\sin(v+\eta)=a(v)
\nonumber\\
W(-1,-1)|_{1,1}&=W(1,1)|_{-1,-1}=\rho'\sin(v-\eta)=b(v)
\nonumber\\
W(-1,1)|_{1,-1}&=W(1,-1)|_{-1,1}=\rho'\sin(2\eta)=c(v)
\label{bw6}
\end{eqnarray}
where the Bethe form of the eigenvectors is known 
to hold \cite{bax2002} for all
eigenvectors and all eigenvalues are computed for all $\eta$ and $N.$ 
On the other hand it was recently discovered \cite{dfm}-\cite{fm3} 
that if the root of unity condition 
\begin{equation}
\gamma=2\eta=m\pi/L
\label{root6}
\end{equation}
holds then  the 6 vertex model has an $sl_2$ loop algebra symmetry group.
In this article we review this infinite dimensional symmetry algebra
of the 6 vertex model at roots of unity  and discuss its relation to the 
qualifying restrictions given above for the solution of the
8 vertex model at (elliptic) roots of unity
\begin{equation}
\eta=mK/L.
\label{root8}
\end{equation}

\section{Loop algebra symmetry of the 6 vertex model}

For generic (irrational) values of $\gamma/\pi$ the spectrum of
eigenvalues of the 6 vertex transfer matrix is
nondegenerate. However, when the root of unity condition (\ref{root6})
holds degenerate multiplets occur  if $N>L.$ 
These multiplets may be described in terms of the operator
\begin{equation}
S^z={1\over 2}\sum_{k=1}^N \sigma_k^z
\end{equation}
which commutes with the transfer matrix of the 6 vertex model. Call
$S^z_{max}$ the maximum value of $S^z$ in the multiplet.
Then in the sector
\begin{equation}
S^z\equiv 0~({\rm mod}L)
\label{sec0}
\end{equation}
the number of degenerate states in the multiplet with the value of
$S^z$ given by
\begin{equation}
S^z=S_{\rm {max}}^z-lL~~{\rm with}~0\leq l \leq 2S^z_{\rm {max}}/L
\end{equation}
is
\begin{equation}
{2S^z_{\rm{max}}/L \atopwithdelims() l}
\end{equation}
and thus the total number of states in the degenerate multiplet is
$2^{2S^z_{\rm{max}}/L}$.
When 
\begin{equation}
S^z\equiv n \neq 0 ({\rm mod} L)
\label{secn}
\end{equation}
there are three types of multiplets with degeneracies
\begin{equation}
{[2S^z_{\rm{max}}/L]+(-1,0,1) \atopwithdelims() l}
~~{\rm with}~0\leq l \leq [2S^z_{\rm {max}}/L]+(-1,0,1)
\end{equation}
where $[x]$ is the greatest integer contained in $x$.

These degenerate
multiplets signal the existence of a symmetry of the system
which is not present in the finite system for $\gamma/\pi$ irrational.
This symmetry algebra was discovered in ref. \cite{dfm} where it was
shown that the operators
\begin{eqnarray}
S^{\pm(L)}&=\sum_{1\leq j_1<\cdots<j_L\leq N}q^{L\sigma^z/2}\otimes
\cdots \otimes q^{L\sigma^z/2}\sigma_{j_1}^\pm\otimes
q^{(L-2)\sigma^z/2}\otimes
\cdots \otimes q^{(L-2)\sigma^z/2}\nonumber\\
& \otimes \sigma_{j_2}^{\pm}\otimes q^{(L-4)\sigma^z/2}\otimes \cdots
\otimes \sigma_{j_L}^{\pm} \otimes q^{-L\sigma^z/2}
\otimes \cdots \otimes q^{-L\sigma^z/2}
\end{eqnarray}
 \begin{eqnarray}
T^{\pm(L)}&=\sum_{1\leq j_1<\cdots<j_L\leq N}q^{-L\sigma^z/2}\otimes
\cdots \otimes q^{-L\sigma^z/2}\sigma_{j_1}^\pm\otimes
q^{-(L-2)\sigma^z/2}\otimes
\cdots \otimes q^{-(L-2)\sigma^z/2}\nonumber\\
& \otimes \sigma_{j_2}^{\pm}\otimes q^{-(L-4)\sigma^z/2}\otimes \cdots
\otimes \sigma_{j_L}^{\pm} \otimes q^{L\sigma^z/2}
\otimes \cdots \otimes q^{L\sigma^z/2}
\end{eqnarray}
with $q=-e^{ i \gamma}$ satisfy in the sector (\ref{sec0}) 
the commutation relations with the 6 vertex transfer matrix
${\bf T}_6(v)$
\begin{equation}
[S^{\pm(L)},{\bf T}_6(v)e^{-iP}]=[T^{\pm (L)},{\bf T}_6(v)e^{-iP}]=0
\end{equation}
where $e^{-iP}$ is the lattice translation operator.
We note that $S^{\pm (L)}=T^{\pm (L)*}.$
The operators $S^{\pm (L)},~T^{\pm (L)}$, and $S^z$ 
satisfy the defining relations of the Chevalley
generators of the loop algebra of $sl_2$ 
\begin{eqnarray}
&[S^{+(L)},T^{+(L)}]=[S^{-(L)},T^{-(L)}]=0\label{e1}\\
&[S^{\pm(L)},S^z]=\pm L S^{\pm (L)},~~[T^{\pm(L)},S^z]
=\pm LT^{\pm(L)}\label{e2}\\
&[S^{+(L)},S^{-(L)}]=[T^{+(L)},T^{-(L)}]=-(-q)^L{2\over L}S^z\label{e3}\\
&S^{+(L)3}T^{-(L)}-3S^{+(L)2}T^{-(L)}S^{+(L)}+
3S^{+(L)}T^{-(L)}S^{+(L)2}-T^{-(L)}S^{+(L)3}=0\label{e4}\\
&S^{-(L)3}T^{+(L)}-3S^{-(L)2}T^{+(L)}S^{-(L)}+
3S^{-(L)}T^{+(L)}S^{-(L)2}-T^{+(L)}S^{-(L)3}=0\label{e5}\\
&T^{+(L)3}S^{-(L)}-3S^{+(L)2}S^{-(L)}T^{+(L)}+
3T^{+(L)}S^{-(L)}T^{+(L)2}-S^{-(L)}T^{+(L)3}=0\label{e6}\\
&T^{-(L)3}S^{+(L)}-3T^{-(L)2}S^{+(L)}T^{-(L)}+
3T^{-(L)}S^{+(L)}T^{-(L)2}-S^{+(L)}T^{-(L)3}=0\label{e7}
\end{eqnarray}

In the sector (\ref{secn}) we further define the operators
\begin{eqnarray}
S^{\pm}&=\sum_{j=1}^Nq^{\sigma^x/2}\otimes \cdots q^{\sigma^z/2}\otimes
\sigma^{\pm}_j \otimes q^{-\sigma^z/2}\otimes  \cdots \otimes
q^{-\sigma^z/2}\\
T^{\pm}&=\sum_{j=1}^Nq^{-\sigma^x/2}\otimes \cdots q^{-\sigma^z/2}\otimes
\sigma^{\pm}_j \otimes q^{\sigma^z/2}\otimes  \cdots \otimes
q^{\sigma^z/2}\\
\end{eqnarray}
and we found  in ref. \cite{dfm} that for even $N$ 
the following eight operators
commute with ${\bf T}_6(v)e^{-iP}$ 
\begin{eqnarray}
&(T^+)^n(S^-)^nS^{-(L)},~~S^{-(L)}(S^-)^{L-n}(T^+)^{L-n},\nonumber\\
&S^{+(L)}(S^+)^n(T^-)^n,~~(T^-)^{L-n}(S^{+})^{L-n}S^{+(L)},\nonumber\\
&T^{+(L)}(T^+)^n(S^-)^n,~~(S^-)^{L-n}(T^+)^{L-n}T^{+(L)},\nonumber\\
&(S^{+})^n(T^-)^nT^{-(L)},~~T^{-(L)}(T^-)^{L-n}(S^+)^{L-n}.
\label{comt6}
\end{eqnarray} 
The  operators
\begin{equation}
(T^+)^n(S^-)^n,~(S^+)^n(T^-)^n,~(T^-)^{L-n}(S^+)^{L-n},~(S^-)^{L-n}(T^+)^{L-n}
\label{proj}
\end{equation}
which appear in (\ref{comt6}) each have a large null space but they
are not in themselves projection operators. For $L=2$  and numerically
on the computer for $L=3$ we have constructed the projection operators
onto the eigenspace of the nonzero eigenvalues of the operators
(\ref{proj}) and using (\ref{comt6}) have constructed the
corresponding projections of $S^{\pm(L)}$ and $T^{\pm(L)}$ and 
have verified that for the projected operators in the sector
(\ref{secn}) that (\ref{e1}),(\ref{e2}) and (\ref{e4})-(\ref{e7}) 
hold without modification but that in (\ref{e3}) the constant $2/L$ is
replaced by an expression which depends on both $L$ and $n.$ We thus
conclude that for even $N$ the $sl_2$ loop algebra is a symmetry algebra of all
sectors of the 6 vertex model.

\section{Evaluation Representations, Drinfeld polynomials and Bethe
Vectors}

The degenerate multiplets of  the 6 vertex model are an example of
a ``highest weight'' phenomenon where all eigenvectors of the multiplet
may be obtained by letting the generators of the symmetry algebra
operate on the ``highest weight vector'' $|\Omega>$ of the multiplet.
In the sector $S^z\equiv 0~({\rm mod}L)$ the highest weight vector
$|\Omega >$ is defined \cite{cp1},\cite{deg1} in terms of the 
Chevalley generators $S^{\pm(L)}$ and $T^{\pm(L)}$ of
the previous section as
\begin{eqnarray}
&S^{+(L)}|\Omega>=T^{+(L)}|\Omega>=0\label{def1}\\
&{T^{+(L)r}\over r!}{S^{-(L)r}\over r!}|\Omega>=\mu_r|\Omega>,
{S^{+(L)r}\over r!}{T^{-(L)r}\over r!}|\Omega>=\mu_r|\Omega>,
\label{def2}\\
&S^z|\Omega>=S^z_{\rm max}|\Omega >\label{def3}.
\end{eqnarray}

To further study these finite dimensional representation we need to
recall a fundamental property of all affine Lie algebras that, besides
the Chevalley basis, they are also characterized by a ``mode'' basis. In
this basis the elements of the $sl_2$ loop algebra are $e(n),~f(n)$
and $h(n),$ where $n$ is an integer, which satisfy the commutation
relations
\begin{eqnarray}
&[e(m),f(n)]=h(m+n)\nonumber\\
&[e(m),h(n)]=-2e(m+n)\nonumber\\
&[f(m),h(n)]=2f(m+n).
\end{eqnarray}
The relation to the Chevalley basis of the previous section is
\begin{equation}
e(0)=T^{-(L)},~~e(-1)=S^{-(L)}~~
f(0)=T^{+(L)},~~f(1)=S^{+(L)}.
\end{equation}
In terms of this mode basis the evaluation representations are
specified by vectors $|a_j,m_j>$ where for all integer $n$ (positive,
negative or zero)
\begin{eqnarray}
e(n)|a_j,m_j>=a_j^ne_{m_j}|a_j,m_j>\nonumber\\
f(n)|a_j,m_j>=a_j^nf_{m_j}|a_j,m_j>\nonumber\\
h(n)|a_j,m_j>=a_j^nh_{m_j}|a_j,m_j>
\end{eqnarray}
where $a_j$ are called evaluation parameters and $e_{m_j},~f_{m_j}$ and
$h_{m_j}$ are a spin $m_j/2$ representation of $sl_2.$ An important
theorem \cite{cp1},\cite{deg1} is that the evaluation 
parameters $a_j$ are the roots with multiplicities $m_j$
of what is
called the (classical) Drinfeld polynomial $P_{\Omega}(z)$ 
\begin{equation}
P_{\Omega}(z)=\sum_{r\geq 0}\mu_r(-z)^r
\label{drin}
\end{equation}
where $\mu_r$ are the eigenvalues defined in (\ref{def2}).

To apply this notion of highest weight vector to the present case we
need to find the relation between $|\Omega >$ and the Bethe form of
the eigenvectors. In the region $S^z\geq 0$ these
Bethe vectors are specified by the coordinate $x_k$ of
$n={N\over 2}-S^z$
``down'' spins which satisfy $1\leq x_1<x_2<\cdots <x_n\leq N$ and the
form of the wave function is
\begin{equation}
|x_1,x_2,\cdots x_n>=\sum_P A_P e^{i(k_{P1}x_1+k_{P2}x_2+\cdots
 +k_{Pn}x_n)}
\label{wf}
\end{equation}
where the sum is over all $n!$ permutations $P$ of $1,\cdots n$, The
$A_P$ are specified functions of the $k_{Pj}$ and the $k_j$ are given
in terms of $v_j$ by
$e^{ik}=(e^{i\gamma}-e^{iv})/(e^{i(v+\gamma)}-1)$
where the $v_j$ satisfy 
\begin{equation}
{\sin(v_j+\gamma/2)\over \sin(v_j-\gamma/2)}=\prod_{l=1,\neq
j}^{N/2-S^z}{\sin(v_j-v_l+\gamma)\over \sin(v_j-v_l-\gamma)}.
\label{beq}
\end{equation}
These equations uniquely specify the eigenvectors as long as the root
of unity condition (\ref{root6}) does not hold. 
In terms of the ``algebraic Bethe Ansatz'' presented in ref.
\cite{tf},\cite{kbi} the states (\ref{wf}) are given as 
$\prod_j B(v_j)|0>$ where the $v_j$ are determined from
(\ref{beq}),$|0>$ is the state with no down spins and the operator $B(v)$
is the upper right hand element of the $2\times 2$ monodromy matrix
${\bf M}(v)$ given as
\begin{equation}
\left(\begin{array}{cc}
A(v)&B(v)\\
C(v)&D(v)
\end{array}\right)=
{\bf M}(v)=W(\mu_1,\nu_1)W(\mu_2,\nu_2)\cdots W(\mu_N,\nu_N)
\end{equation}

When the root of unity condition (\ref{root6}) does hold there is
ambiguity in the solution of (\ref{beq}) because $L$ of the $v_j$ 
may be of the form
\begin{equation}
v_{j;k}=v^c_j+km\pi/L,~~k=0,1,\cdots,L-1.
\label{6lstrings}
\end{equation}
These sets of roots are called ``complete strings'' and give factors
of $0/0$ which cancel  out of (\ref{beq}).
It is indirectly shown in ref.\cite{fm3} that the highest weight vectors
$|\Omega>$ are all Bethe vectors where the ``Bethe roots'' $v_j$ do
not contain any complete $L$ strings. The remaining members of the
multiplet, while they are still of the Bethe form (\ref{wf}), will contain
complete $L$ strings. 
For $\gamma$ at roots of unity the algebraic Bethe Ansatz does not give a proper construction of eigenstates 
belonging to degenerate multiplets as the operator $\prod_{k=0}^{L-1}B(v-2ik\gamma)$ which formally
creates a $L$-string vanishes \cite{tar}. We have shown in \cite{fm3} 
that the creation operator of the $L$-string part of a state vector is
\begin{eqnarray}
&&B^{(L)}(v)= \nonumber \\
&&\sum_{k=0}^{L-1}\left( \prod_{l=0}^{k-1} B(v-2il\gamma)
\right)\left( B_{\gamma}(v-2ik\gamma)+{X(v-2ik\gamma)\over
Y(v)}B_v(v-2ik\gamma)\right)\times\left(\prod_{l=k+1}^{L-1}B(v-2il\gamma)\right)
\label{bcurrent}
\end{eqnarray}
where $B_{\gamma}(v)$ and $B_{v}(v)$ specify derivatives of $B(v)$
with respect to $\gamma$ and $v$ respectively and where
\begin{eqnarray}
X(v)&&=2i\sum_{l=0}^{L-1}\frac{l \sinh^N{1\over 2}(v-(2l+1)i\gamma)}
{\prod_{k=1}^n \sinh{1\over 2}(v-v_k-2il\gamma)\sinh{1\over
2}(v-v_k-2i(l+1)\gamma)}
\label{xres}
\end{eqnarray}
and 
\begin{equation}
Y(v)=\sum_{l=0}^{L-1}\frac{\sinh^N{1\over 2}(v-(2l+1)i\gamma)}
{\prod_{k=1}^n \sinh{1\over 2}(v-v_k-2il\gamma)
\sinh{1\over 2}(v-v_k-2i(l+1)\gamma)}
\label{yres}
\end{equation}
and $v_k$ with $k=1,\cdots , n$ are the ordinary Bethe roots.
We show in \cite{fm3} that $Y(v)$ satisfies the periodicity condition 
\begin{equation}
Y(v+m\pi/L)=Y(v)
\end{equation}
and thus is a Laurent polynomial in
$z=e^{2iLv}$.
Furthermore we define the degrees $d_{\pm}$ by
\begin{equation}
Y(v)=C_{\pm}e^{\pm 2iL d_{\pm}v}~~{\rm as}~v\rightarrow \pm
i\infty.
\end{equation}
The Drinfeld polynomial $P_{\Omega}(z)$ is then  given as \cite{fm3}
\begin{equation}
P_{\Omega}(z)=e^{d_{-}2 iLv}Y(v)
\label{drin2}
\end{equation}
which is a polynomial in $z$ of degree $d=d_{+}+d_{-}.$

If the zeros of $Y(v)$ all have multiplicity one then the number of
eigenvalues in the multiplet specified by the evaluation parameters 
of the highest weight vector $|\Omega>$ is $2^d.$ There exists no
analytic proof that the roots of $Y(v)$ are all simple but this has
been verified in all numerical evaluations which have been made.

\section{The 8 vertex model at roots of unity}

The eigenvalues of the 8 vertex model develop degenerate multiplets
when $\eta$ satisfies (\ref{root8}) in a manner almost identical with
the degeneracies of the 6 vertex model at (\ref{root6}). The only
difference is that some of the 6 vertex multiplets split into two
multiplets for the 8 vertex model when the elliptic nome $p\neq 0.$
Thus the numerical evidence for the existence of a symmetry algebra
for the 8 vertex model at the roots of unity points (\ref{root8}) is
just as compelling as for the 6 vertex model.

However, despite the great similarity in the numerical evidence the
treatment of the degeneracies of the 8 vertex model is very different
from what was done for the 6 vertex model. In particular we do not
have expressions for operators like $S^{\pm (L)}$ and $T^{\pm (L)}$ which
commute with ${\bf T}_8e^{-iP}$ and we do not know what is the symmetry
algebra which produces the degeneracy. Accordingly we cannot give a
group theoretic explanation in terms of highest weight states and
evaluation parameters, 

Instead of using group theory we study the multiplets of the 8 vertex
model at roots of unity (\ref{root8}) by use of the matrix 
first introduced by Baxter \cite{bax72} in 1972 in his original
solution of the 8 vertex model. This matrix (which we call ${\bf
Q}_{72}(v)$)
satisfies the functional equation 
\begin{equation}
{\bf T}_8(v){\bf Q}(v)=[\rho \Theta(0)h(v-\eta)]^N{\bf Q}(v+2\eta)
+[\rho\Theta(0)h(v+\eta)]^N{\bf Q}(v-2\eta)
\label{tq}
\end{equation}
where
$h(v)=\Theta(v)H(v)$
and the commutation relations
\begin{equation}
[{\bf T}_8(v), {\bf Q}(v')]=[{\bf Q}(v),{\bf Q}(v')]=0.
\label{tqqq}
\end{equation}
In Baxter's paper \cite{bax72} the following explicit form is given for
${\bf Q}_{72}(v)$
\begin{equation}
{\bf Q}_{72}(v)={\bf Q}_R(v){\bf Q}_{R}^{-1}(v_0)
\end{equation} 
where $v_0$ is an arbitrary normalization point at which ${\bf Q}_{R}(v)$
is nonsingular. The matrix ${\bf Q}_{R}(v)$ is defined as
\begin{equation}
{\bf Q}_R(v)|_{\alpha |\beta}={\rm Tr}{\bf S}(\alpha_1, \beta_1)
{\bf S}(\alpha_2, \beta_2)\cdots {\bf S}(\alpha_N, \beta_N)
\end{equation}
where $\alpha_j$ and $\beta_j=\pm 1$ and $S(\alpha,\beta)$ is an
$L\times L$ matrix given as (C16) of ref.\cite{bax72}
\begin{equation}
{\bf S}(\alpha,\beta)=\left( \begin{array}{cccccc}
z_0 & z_{-1}&0&0&\cdot&0\\
z_1 & 0  &z_{-2}&0&\cdot&0\\ 
0   &z_2 &0&z_{-3}&\cdot&0\\
\cdot&\cdot&\cdot&\cdot&\cdot&\cdot\\
0&0&0&\cdot&0&z_{1-L}\\
0&0&0&\cdot&z_{L-1}&z_L
\end{array}\right)
\end{equation}
with 
$z_m=q(\alpha, \beta, m|v)$
and 
\begin{eqnarray}
q(+,\beta,m|v)&=&H(v+K+2m\eta)\tau_{\beta,m},\nonumber\\
q(-,\beta,m|v)&=&\Theta(v+K+2m\eta)\tau_{\beta,m}
\label{qtau}
\end{eqnarray} 
The $\tau_{\beta,m}$ are generically arbitrary but we note that if
they are all set equal to unity then ${\bf Q}_R(v)$ is so singular that
its rank becomes 1. On the other hand as long as the $\tau_{\beta,m}$
are chosen so that there is a $v_0$ such that ${\bf Q}_R(v_0)$ is not singular
then ${\bf Q}_{72}(v)$ is independent of $\tau_{\beta,m}$. 

For many years it seems to have been assumed that for all $L$ and all
$N$  there did exist a $v_0$ for which ${\bf Q}_R(v)$ was 
nonsingular but in  ref. \cite{fm4} we made  a computer study  
of ${\bf Q}_{72}(v)$ which demonstrated that such a general statement of
nonsingularity does not hold. Instead we found that while ${\bf
Q}_{R}(v)$ is nonsingular if $L$ is even or when $L$ and $m$ are odd
that if $L$ is odd and $m$ is even then ${\bf Q}_R(v)$ is singular
for all $v$ when $N$ is odd or when $N$ is even and $L\geq N-1.$ 
When $N$ is even the eigenvalues of ${\bf T}_8(v)$ may still be
studied by means of the $TQ$ equation (\ref{tq}) even though 
${\bf Q}_{R}(v)$ is singular for $L\geq N-1$ by use of the symmetry
\begin{equation}
{\bf T}_8(v+K;K-\eta)={\bf S}{\bf T}_8(v;\eta)
\label{8sym}
\end{equation}
where
\begin{equation}
{\bf S}=\prod_{j=1}^N\sigma_j^z
\end{equation}
and we note that
\begin{equation}
~[{\bf S},{\bf T}_8(v)]=0,~~
~[{\bf S},{\bf Q}_{72}(v)]=0\label{sq}.
\end{equation}
However for odd $N$ with $L$ odd and $m$ even no such symmetry exists.

Even though ${\bf T}_8(v)$ has many degenerate eigenvalues when $\eta$
satisfies the root of unity condition (\ref{root8}) the matrix 
${\bf Q}_{72}(v)$ has the remarkable property (discovered numerically
but never proven analytically) that it has no degenerate eigenvalues.
Therefore if we can find a criterion to determine the class of
eigenvalues of ${\bf Q}_{72}(v)$ which have the same eigenvectors of
the degenerate eigenvalues of ${\bf T}_8(v)$ we may determine the
degeneracy of an eigenvalue of ${\bf T}_8(v)$ by counting the
corresponding eigenvalues of ${\bf Q}_{72}(v)$. To determine this relation
between eigenvalues of ${\bf T}_8(v)$ and ${\bf Q}_{72}(v)$ 
we note, as demonstrated in ref. \cite{fm4}, that ${\bf Q}_{72}(v)$
obeys the following quasiperiodicity conditions
\begin{eqnarray}
&{\bf Q}_{72}(v+2K)={\bf S}{\bf Q}_{72}(v)\label{mqp1}\\
&{\bf Q}_{72}(v+2iK')=p^{-N}{\rm exp}(-iN\pi v/K){\bf Q}_{72}(v).\label{mqp2}
\end{eqnarray}
It follows from (\ref{sq}) that ${\bf S}$ and ${\bf Q}_{72}(v)$ may be
simultaneously diagonalized and thus in the  basis where ${\bf S}$ is
diagonal with eigenvalues $(-1)^{\nu'}$ with $\nu '=0,1$ we obtain
from (\ref{mqp1}),(\ref{mqp2})  the
quasiperiodicity conditions for the eigenvalues 
\begin{eqnarray}
&Q_{72}(v+2K)=(-1)^{\nu'}Q_{72}(v)\label{qp1}\\
&Q_{72}(v+2iK')=p^{-N}{\rm exp}(-iN\pi v/K)Q_{72}(v).\label{qp2}
\end{eqnarray}
and thus the eigenvalues $Q_{72}(v)$ of the matrix ${\bf Q}_{72}(v)$ 
may be expressed in a factored form as
\begin{equation}
Q_{72}(v)={\cal K}(p;v_k){\rm exp}(-i\nu\pi v/2K)\prod_{j=1}^NH(v-v_j)
\label{qfac}
\end{equation}
where we have the sum rules
\begin{eqnarray}
\nu=\sum_{j=1}^N{\rm Im}v_j/K'={\rm even~integer}~-\nu'=N\\
N+\sum_{j=1}^N {\rm Re}v_j/K={\rm even~integer}.
\end{eqnarray}

From the commutation relations (\ref{tqqq}) we see that all matrices in
the $TQ$ equation (\ref{tq}) may be simultaneously diagonalized 
and making this diagonalization we obtain an equation for eigenvalues 
$T_8(v)$ and $Q(v)$
\begin{equation}
{ T}_8(v){ Q}(v)=[\rho \Theta(0)h(v-\eta)]^N{ Q}(v+2\eta)
+[\rho\Theta(0)h(v+\eta){Q}(v-2\eta)
\label{evtq}
\end{equation}
Then using the factored form (\ref{qfac}) of the eigenvalues
$Q_{72}(v)$ in (\ref{evtq}) 
and setting $v=v_j$   
we obtain the equation for the zeros $v_j$ of $Q_{72}(v)$
\begin{equation}
\left({h(v_l-mK/L)\over h(v_l+mK/L)}\right)^N=e^{2\pi i \nu
m/L}\prod_{j=1,\neq l}^N{H(v_l-v_j-2mK/L)\over H(v_l-v_j+2mK/L)}. 
\label{8geq}
\end{equation}
 
If in the complete set of $N$ roots $v_j$ there are sets of $L$ roots
\begin{equation}
v_{j;k}=v^c_j+2kmK/L
\label{8string}
\end{equation}
we see from the eigenvalue expression (\ref{evtq}) that all terms with
$v^c_j$ cancel and thus eigenvalues of $Q_{72}(v)$ which differ only  
in the location of the string centers $v^c_j$ have the same degenerate
eigenvalues $T_8(v).$ Thus we can count the degeneracy of an
eigenvalue of ${\bf T}_8(v)$ by determining all those eigenvalues 
$Q_{72}(v)$ which differ only by their $L$ strings. 
These $L$ string
solutions are the analogue for the 8 vertex model of the $L$ string
solutions (\ref{6lstrings}) of the 6 vertex model and like the 
$L$ strings of the 6 vertex model the string centers cannot be
determined from the equation (\ref{8beq}).
We note that the strings (\ref{8string}) which contain $L$ roots are
not the same as the strings of ref. \cite{bax733},\cite{deg2} 
which contain $2L$ roots and are invariant under
translation by $iK'$.

Unlike the 6 vertex equations (\ref{beq})  
which have $N/2-S^z$ Bethe roots $v_j$ the 8 vertex equations
(\ref{8geq}) have $N$ roots for all eigenstates of ${\bf T}_8(v).$
Moreover there are distinct features in the solutions  $v_j$ of
(\ref{8geq}) which depend on $N$ which do not occur for
(\ref{beq}). To see these features it is (at present) necessary to do
a numerical study of the zeroes of the eigenvalues of ${\bf Q}_{72}(v).$
We have done this for $N$ even in ref. \cite{fm4} and for $N$ odd in
ref. \cite{fm5} and have found the following results for the roots of
unity condition (\ref{root8}). 

\vspace{.1in}

{ \bf Even $N$ with $m$ odd and $L$ even or odd}

There are $n_B$ pairs of roots 
\begin{equation}
v_j^N,~~v_j^B+iK'
\label{pair}
\end{equation}
which we call Bethe roots and $n_L$ complete $L$ strings of the form
(\ref{8string})
where 
\begin{equation}
2n_B+Ln_L=N
\label{bls}
\end{equation}
When this form is used in the $TQ$ equation (\ref{8beq}) we find that
the $n_B$ Bethe roots satisfy
\begin{equation}
\left({h(v^B_l-mK/L)\over h(v^B_l+mK/L)}\right)^N=e^{2\pi i (\nu-n_B)
m/L}\prod_{j=1,\neq l}^{n_B}{h(v^B_l-v_j-2mK/L)\over h(v^B_l-v_j+2mK/L)}. 
\label{8beq}
\end{equation}
The $L$ string roots are not determined from this equation and 
we proved for $L=2$ in ref.\cite{fusion} and conjectured for $L\geq3$ in ref.\cite{fm4}
that they are determined in terms of the
$n_B$ Bethe roots by the functional equation
\begin{equation}
{\bf A}'e^{-N\pi i v/2K}{\bf Q}_{72}(v-iK')
=\sum_{l=0}^{L-1}{h^N(v-(2l+1)\eta){\bf Q}_{72}(v)\over
{\bf Q}_{72}(v-2l\eta){\bf Q}_{72}(v-2(l+1)\eta)}
\label{conj}
\end{equation}
where ${\bf A}'$ is a matrix 
which commutes with ${\bf Q}_{72}(v),$ 
is independent of $v$ and depends on the
normalization in the construction of ${\bf Q}_{72}(v).$ 
The left hand side of (\ref{conj}) is an entire function and thus the
apparent poles on the right hand side at the zeroes of 
${\bf Q}_{72}(v)$ must cancel. This cancellation leads to the equation
for the Bethe roots $v^B_j$ of (\ref{8beq}). The remaining zeroes of
the right hand side give the string solutions $v^L_{j;k}$ and have the
property that if $v^L_{j;k}=v^c_j+2kmK/L$ is a solution then
$v^L_{j;k}+iK'$ is also a solution. Therefore the number of
eigenvalues of ${\bf Q}_{72}(v)$ which correspond to a degenerate
eigenvalue of ${\bf T}_8(v)$ is $2^{n_L}.$ Thus for even $N$ 
we have explained the
degeneracies of the 8 vertex model transfer matrix without an appeal
to group theory.

\vspace{.1in}

{\bf Odd $N$ with $m$ odd and $L$ even or odd} 

There are no paired roots and no $L$ strings. All the $N$ roots $v_j$ 
are determined from (\ref{8geq}). For every set of roots $v_j$ which
solves (\ref{8geq}) there is a second solution $v_j+iK'$ which also
solves (\ref{8geq}) and thus all eigenvalues of ${\bf T}_8(v)$
are doubly degenerate. This is
to be expected from the fact that the transfer matrix is invariant
under spin reversal and that all states have half integer total spin.
Because there are never any $L$ strings there are no further
degeneracies in the eigenvalue spectrum of ${\bf T}_8(v)$.

\vspace{.1in}

There remains the case $L$ odd and $m$ even where ${\bf Q}_{72}$ does
not exist. For $N$ even this is
related to the case $m$ odd  by use of the symmetry (\ref{8sym}). 
For $N$ odd, there is no such symmetry. However, it was demonstrated
in ref. \cite{bax731}-\cite{bax733} that if there exist integer $n_B$
and $n_L$ such that (\ref{bls}) holds that the eigenvalues of the
transfer matrix ${\bf T}_8(v)$ may be computed. Because $N$ and $L$
are odd this requires that $n_L$ be odd (and in particular non zero)
for the method to apply.    

There is no proof that for $N$ odd, $L$ odd and $m$ even that
there is any matrix ${\bf Q}(v)$ which satisfies
the $TQ$ equation (\ref{tq}) and the commutation relations
(\ref{tqqq}). Nevertheless we conjecture that such a matrix 
${\bf Q}(v)$ does exist which satisfies the quasiperiodicity conditions
(\ref{qp1}) and (\ref{qp2}). We have studied this numerically in
ref.\cite{fm5} and found that  
for $N$ odd, with $L$ odd and $m$ even there are two types of
eigenvalues $Q(v)$ which occur:

\vspace{.1in}

Type I.

There are $n_B$ pairs of Bethe roots (\ref{pair}) and $n_L$ complete $L$
strings of the form (\ref{8string}) where (\ref{bls}) holds with
$n_L\neq 0.$ The Bethe roots are determined from (\ref{8beq}) and the
method of ref. \cite{bax731}-\cite{bax733} applies.
As with the case of even $N$ we find that for each complete $L$ string
$v^L_{j;k}$ there is a companion string $v^L_{j;k}+iK'.$ Therefore the
degeneracy of the transfer matrix eigenvalue is $2^{n_L}.$ 
 
\vspace{.1in}

Type II.

There are no pairs of Bethe roots and no $L$ strings. To every set of
roots $v_j$ there is a companion set of roots $v_j+iK'$ and thus all
eigenvalue of the transfer matrix ${\bf  T}_8(v)$ are doubly
degenerate. Because there are no paired roots and no complete $L$
string the condition (\ref{bls}) does not hold and the method of
computing eigenvectors of ref. \cite{bax731}-\cite{bax733} does not
apply to these states. On the other hand we have verified that the
function $Q(v)$ constructed from $v_k$ according to (\ref{qfac}) 
satisfies the equation (\ref{conj}) with $A'=0$ 
from which (\ref{8geq}) follows. This case for which the methods of
ref. \cite{bax731}-\cite{bax733} fail
is particularly interesting. A
special case was first studied in ref. \cite{bax89} and the 6 vertex limit
has been studied extensively by several authors \cite{strog}-\cite{korff}.

\section{Outlook}

The results and computations presented above provide a detailed
explanation of the degeneracies in the eigenvalue spectrum of the 6 and
8 vertex models at roots of unity. However it is clear that there are
many open questions which need to be resolved before the problem of the
degeneracies can be considered to be solved. 

For the 6 vertex model with $S^z\equiv 0~({\rm mod}L)$ where the
generators of the symmetry algebra are explicitly known the highest
weight property (\ref{def1})-({\ref{def3}) of the Bethe vectors 
which contain no strings
needs to be directly proven \cite{deg1}. For $S^z\neq 0~({\rm mod}L)$ the projection
operators needed for the several different sectors need to be studied
in such a form that the symmetry algebra in the sectors may be
analytically established. For all sectors a direct proof of the
expression for the Drinfeld polynomial (\ref{drin2}) 
in terms of the Bethe roots
needs to be given and a proof that the roots of the Drinfeld
polynomial are all simple needs to be found.
In addition it would be desirable to find a physical interpretation for
the basis of degenerate eigenvectors which is specified by the
evaluation parameters.

For the 8 vertex model much more needs to be done because here the
symmetry algebra is not known even though such an algebra must exist.
This algebra for the 8 vertex model must contain information about the
sectoring of the 6 vertex model for $S^z\neq 0~({\rm mod}L).$
There should presumably be some analogue for this symmetry algebra of
the highest weight phenomenon and the righthand side of
(\ref{conj}) should be some sort of elliptic generalization of a
Drinfeld polynomial with the zeroes providing a generalization of
the evaluation parameters.

Moreover, while for the 6 vertex model the eigenvectors and
eigenvalues of the transfer matrix $T_6(v)$ are known for all $\gamma$
and all $N$ to follow from the Bethe form of the eigenvectors the same
is not true for the 8 vertex model. For generic $\eta$ and even $N$ 
a matrix ${\bf Q}_{73}(v)$ is known
\cite{bax731}-\cite{bax733},\cite{baxb} which satisfies the $TQ$
equation (\ref{tq}) and the commutation relations (\ref{tqqq}) but
this matrix does not specialize to ${\bf Q}_{72}(v)$ when $\eta$ is
root of unity. This demonstrates that the matrix ${\bf Q}(v)$ which
satisfies (\ref{tq}) and (\ref{tqqq}) is not unique and it is of
interest to find how many arbitrary parameters can be contained in the
solutions. This would generalize the studies of the ${\bf Q}(v)$
matrices 
made for the 6 vertex model \cite{bs}-\cite{korff4}.
Moreover for odd $N$ in the case $L$ odd with $m$ even
or for generic $\eta$ no ${\bf Q}(v)$ matrix is proven to exist even though
we have numerically seen that the $TQ$ equation can be satisfied.
The case $N$ odd $L$ odd and $m$ even
needs to be investigated in the detail for which the 6 vertex limit has been
studied in ref. \cite{strog}-\cite{korff}. 
The recent work of Bazhanov and Mangazeev \cite{bm} for
$\eta=2K/3$ is an important development in this study.
\vspace{.1in}

{\large \bf Acknowledgments}

\vspace{.1in}

One of us (BMM) is pleased to thank Prof. M. Kashiwara
for hospitality at the Research Institute of Mathematical
Sciences of Kyoto University where part of this work was carried out,
to M. Shiraishi for the invitation to participate in this workshop
and to T. Deguchi, M. Jimbo, C. Korff and T. Miwa for many useful discussions.
This work is partially supported by NSF grant DMR0302758.

\appendix

\section{Properties of theta functions}

The definition of Jacobi theta functions of nome $p$ is
\begin{equation}
H(v)=2\sum_{n=1}^{\infty}(-1)^{n-1}p^{(n-{1\over 2})^2}
\sin [(2n-1)\pi v/2K]
\label{hdef}
\end{equation}
\begin{equation}
\Theta(v)=1+2\sum_{n=1}^{\infty}(-1)^np^{n^2}\cos (nv\pi/K)
\label{thetadef}
\end{equation}
where $K$ and $K'$ are the standard elliptic integrals of the first kind and
\begin{equation}
p=e^{-\pi K'/K}
\end{equation}
These theta functions satisfy the quasiperiodicity relations
\begin{eqnarray}
H(v+2K)&=-H(v)\label{p1}\\
H(v+2iK')&=-p^{-1}e^{-\pi i v/K}H(v)\label{p2}\\
\Theta(v+2K)&=\Theta(v)\label{p3}\\
\Theta(v+2iK')&=-p^{-1}e^{-\pi i v/K}\Theta(v)\label{p4}
\end{eqnarray}
From (\ref{hdef}) and (\ref{thetadef}) we see that $\Theta(v)$ and $H(v)$
are not independent but satisfy
\begin{eqnarray}
\Theta(v\pm iK')&=\pm ip^{-1/4}e^{\mp{\pi i v\over 2 K}}H(v)\nonumber\\
H(v\pm iK')&=\pm ip^{-1/4}e^{\mp{\pi i v\over 2 K}}\Theta(v)
\label{thtrans}
\end{eqnarray}

\end{document}